\documentclass[prl,twocolumn,amsmath,amssymb,longbibliography]{revtex4-1}

\usepackage{hyperref}
\usepackage{graphicx}

\usepackage{amsmath}
\usepackage{bm}
\usepackage{amssymb}
\usepackage{ulem}

\usepackage{color}

\newcommand{\tr}[1]{{\textrm{#1}}}

\begin{document}

\title{Quantum confined Stark effect in a MoS$_2$ monolayer van der Waals heterostructure}

\author{Jonas G. Roch}
\email{jonasgael.roch@unibas.ch}
\affiliation{Department of Physics, University of Basel, Klingelbergstrasse 82, CH-4056 Basel, Switzerland}

\author{Nadine Leisgang}
\affiliation{Department of Physics, University of Basel, Klingelbergstrasse 82, CH-4056 Basel, Switzerland}

\author{Guillaume Froehlicher}
\email{g.froehlicher@unibas.ch}
\affiliation{Department of Physics, University of Basel, Klingelbergstrasse 82, CH-4056 Basel, Switzerland}

\author{Peter Makk}
\affiliation{Department of Physics, University of Basel, Klingelbergstrasse 82, CH-4056 Basel, Switzerland}
 
\author{Kenji Watanabe}
\affiliation{National Institute for Material Science, 1-1 Namiki, Tsukuba, 305-0044, Japan}

\author{Takashi Taniguchi}
\affiliation{National Institute for Material Science, 1-1 Namiki, Tsukuba, 305-0044, Japan}

\author{Christian Sch\"{o}nenberger}
\affiliation{Department of Physics, University of Basel, Klingelbergstrasse 82, CH-4056 Basel, Switzerland}

\author{Richard J. Warburton}
%\email{richard.warburton@unibas.ch}
\affiliation{Department of Physics, University of Basel, Klingelbergstrasse 82, CH-4056 Basel, Switzerland}


\begin{abstract}
The optics of dangling-bond-free van der Waals heterostructures containing transition metal dichalcogenides are dominated by excitons. A crucial property of a confined exciton is the quantum confined Stark effect (QCSE). Here, such a heterostructure is used to probe the QCSE by applying a uniform vertical electric field across a molybdenum disulfide (MoS$_2$) monolayer. The photoluminescence emission energies of the neutral and charged excitons shift quadratically with the applied electric field provided the electron density remains constant, demonstrating that the exciton can be polarized. Stark shifts corresponding to about half the homogeneous linewidth were achieved. Neutral and charged exciton polarizabilities of $(7.8~\pm~1.0)\times 10^{-10}~\tr{D~m~V}^{-1}$ and $(6.4~\pm~0.9)\times 10^{-10}~\tr{D~m~V}^{-1}$ at relatively low electron density ($8 \times 10^{11}~\tr{cm}^{-2}$) have been extracted, respectively. These values are one order of magnitude lower than the previously reported values, but in line with theoretical calculations. The methodology presented here is versatile and can be applied to other semiconducting layered materials as well.
\end{abstract}

%\begin{tocentry}
%    \begin{center}
%    \includegraphics[width=8cm]{TOC.pdf}
%    \end{center}
%\end{tocentry}

\maketitle
 
The recent emergence of optically-active layered semiconductors~\cite{Mak2010,Splendiani2010}, such as molybdenum disulfide (MoS$_2$), and of the so-called van der Waals heterostructures (vdWhs)~\cite{Geim2013,Novoselov2016} pave the way towards engineered quantum structures. Excitons in MoS$_2$ and other transition metal dichalcogenides have particularly large exciton binding energies~\cite{Chernikov2014} such that excitons dominate the optical properties, even at room temperature. Therefore, the fundamental properties of the excitons need to be elucidated. A basic feature of semiconductor nanostructures is the quantum confined Stark effect (QCSE), the change in optical response on applying an electric field perpendicular to the layers~\cite{Miller1984}. On the one hand, the QCSE characterizes the sensitivity of the exciton energy to charge noise as charge noise results in a fluctuating electric field within the device. The QCSE is therefore important in optimizing and understanding optical linewidths. On the other hand, the QCSE can be exploited to trap and manipulate excitons on the nano-scale by applying a locally varying vertical electric field~\cite{High2008,Schinner2013}.

When a DC electric field is applied perpendicular to a MoS$_2$ monolayer ($z$-axis), electrons and holes will tend to move apart in order to decrease their electrostatic potential energy. The resulting energy shift $\Delta E$ of the exciton energy is known as the QCSE and is given by $\Delta E=-\mu_z F_z-\beta_z F_z^2$ where $F_z$ is the component of the electric field, $\mu_z$ the excitonic dipole moment and $\beta_z$ the excitonic polarizability along the $z$-direction. Owing to the reflection symmetry about the molybdenum plane, $\mu_z=0$ in a MoS$_2$ monolayer embedded in a symmetric dielectric environment~\cite{Schuller2013} such that the QCSE is expected to be quadratic in $F_z$.

\begin{figure}[!t]
\begin{center}
\includegraphics[width=8.6cm]{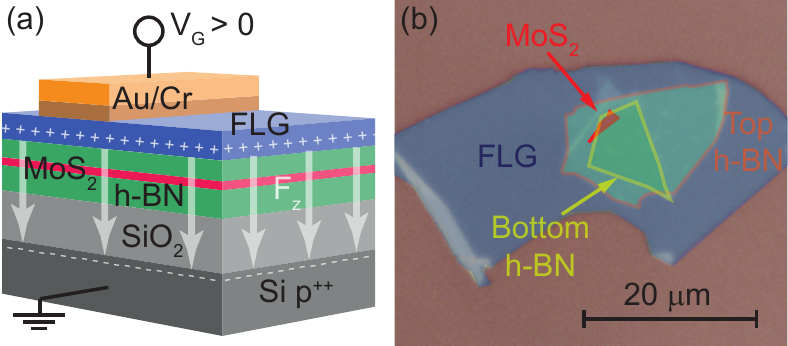}
\caption{(a) Three-dimensional schematic view of the device used to measure the quantum confined Stark effect. The device consists of a MoS$_2$ monolayer sandwiched between two layers of h-BN and covered by a few-layer graphene (FLG) top electrode, deposited onto a highly $p$-doped Si/SiO$_2$ substrate. A voltage $V_\tr{G}$ is applied between the Si substrate and the top electrode to create a uniform electric field across the MoS$_2$. (b) Optical micrograph of the device. The different layers have been artificially highlighted with colors. Photoluminescence is carried out on the part of the MoS$_2$ monolayer which is fully encapsulated in h-BN.}
\label{Fig1}
\end{center}
\end{figure}

\begin{figure*}[!t]
\begin{center}
\includegraphics[width=17.8cm]{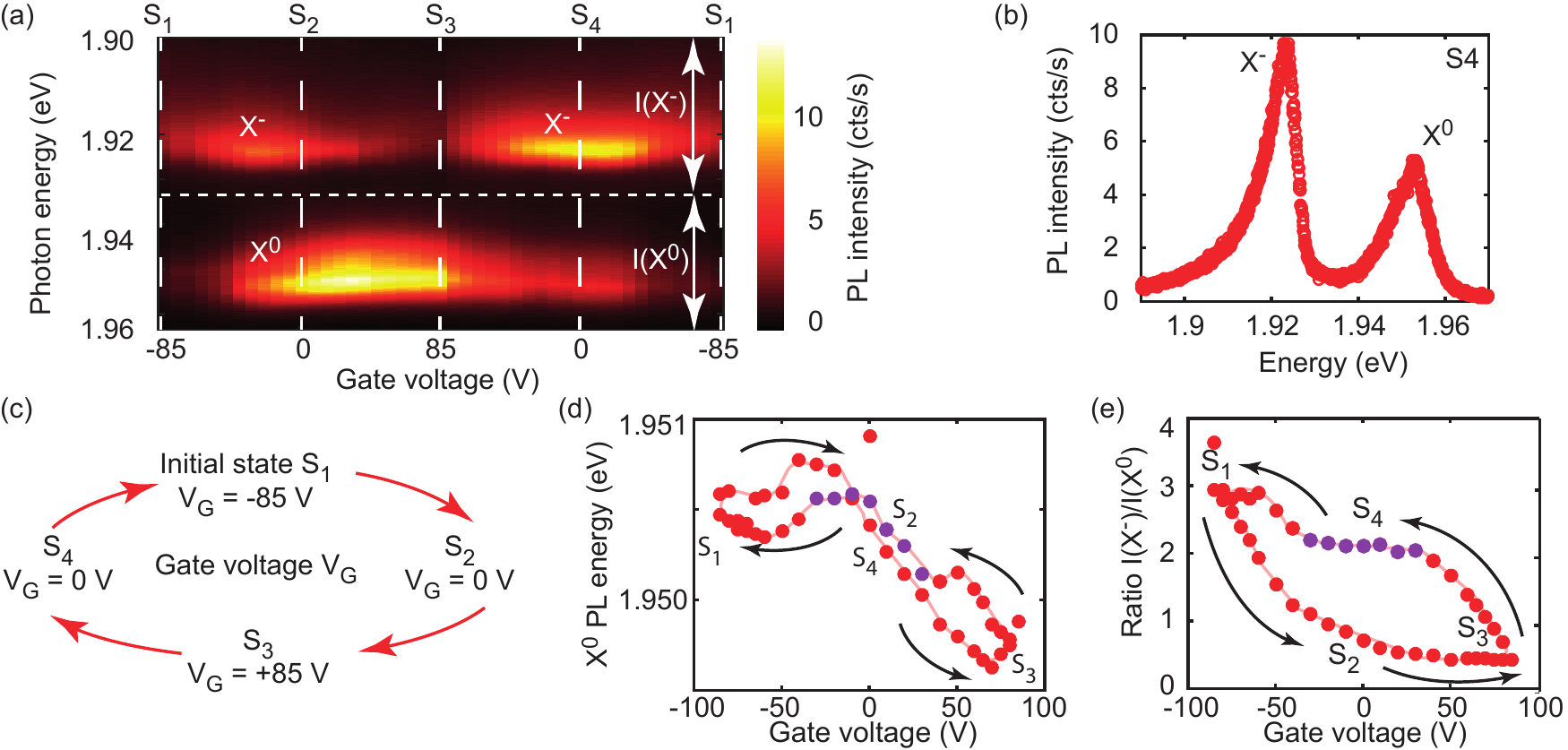}
\caption{(a) Colormap of the photoluminescence spectra of MoS$_2$ as a function of $V_\tr{G}$. The neutral X$^0$ and negatively charged X$^-$ excitons and the four states defined in (c) are labeled. (b) Typical photoluminescence spectrum recorded at S$_4$. (c) The gate voltage $V_\tr{G}$ in the device was varied along a loop from state S$_1$ ($-85$ V) to S$_3$ ($+85$ V) and back, reaching $V_\tr{G}=0$ twice (states S$_2$ and S$_4$). (d) Emission energy of X$^0$ as a function of $V_\tr{G}$. The experimental data points extracted from the spectra in (b) are represented by circles where the purple circles correspond to the data used to measure the Stark shift. The solid line is a guide to the eye and the black arrows indicate the changes made to $V_\tr{G}$. (e) Ratio between the integrated intensity of the X$^-$ and X$^0$ features extracted from the data in (a) as a function of $V_\tr{G}$. The range of integration is indicated by the white double-headed arrows in (a).}
\label{Fig2}
\end{center}
\end{figure*}

Measurement of the Stark shift of the A-exciton in a MoS$_2$ monolayer has been reported~\cite{Klein2016}. However, the experiment was performed on monolayers encapsulated in standard oxides (aluminium and silicon oxides) which have poor optical quality and, most probably, contain a significant density of charge traps~\cite{Guo2015_apl}. Lately, a theoretical study~\cite{Pedersen2016} has predicted $\beta_z$ to be more than one order of magnitude below the reported experimental value. An unambiguous measure of the QCSE in MoS$_2$ is therefore missing. A particular challenge is that the exciton energies depend strongly on the electron density in the MoS$_2$ monolayer~\cite{Chernikov2015}. Furthermore, the description of the optical excitations in the high-density regime has a strong many-electron flavor: the quasi-particles are no longer the simple excitons~\cite{Sidler2016,Efimkin2017}. These considerations mean that the QCSE should be measured at a low and constant electron density.
 
In this Letter, high quality MoS$_2$ monolayers, obtained by encapsulation in hexagonal boron nitride (h-BN), are used to determine precisely the QCSE of the neutral X$^{0}$ and negatively charged X$^{-}$ A-excitons. The photoluminescence (PL) spectra of these samples show narrow linewidths  ($\approx$ 8 meV), close to the ideal limit ($1-2~\tr{meV}$~\cite{Cadiz2017}), allowing the X$^{0}$ and X$^{-}$ to be identified unambiguously. Both spectral features shift when applying an electric field. However, at the same time, the ratio between the integrated intensities of X$^{-}$ and X$^{0}$ varies. The change in this ratio signifies a change in the electron density which, in turn, shifts the emission energies. To separate carefully QCSE and doping contributions to the energy shifts, additional measurements were performed on a directly contacted MoS$_2$ device. These measurements quantify precisely both the X$^{-}$ to X$^{0}$ intensity ratio and the exciton energy shifts as a function of the electron density. We use this information to find a region in the encapsulated device where the electric field can be changed at a constant and relatively low electron density. In this region, we demonstrate a clear QCSE. We determine excitonic polarizabilities typically one order of magnitude smaller than the values reported in Ref.~\onlinecite{Klein2016} but in good agreement with calculations in Ref.~\onlinecite{Pedersen2016}.

The QCSE was measured using the encapsulated device with geometry as depicted in Fig.~\ref{Fig1}(a): two thick h-BN layers are used as dielectric spacers and the top few-layer graphene (FLG) acts as a transparent electrode (see Methods for a description of the fabrication process). Applying a DC voltage $V_\tr{G}$ between the FLG and the highly doped bottom Si substrate creates a uniform electric field in the MoS$_2$ monolayer, oriented perpendicular to the basal plane of the sample. PL spectra were recorded at 4~K as a function of $V_\tr{G}$ in a home-built confocal microscope (see Methods).  

\begin{figure*}[!t]
\begin{center}
\includegraphics[width=17.8cm]{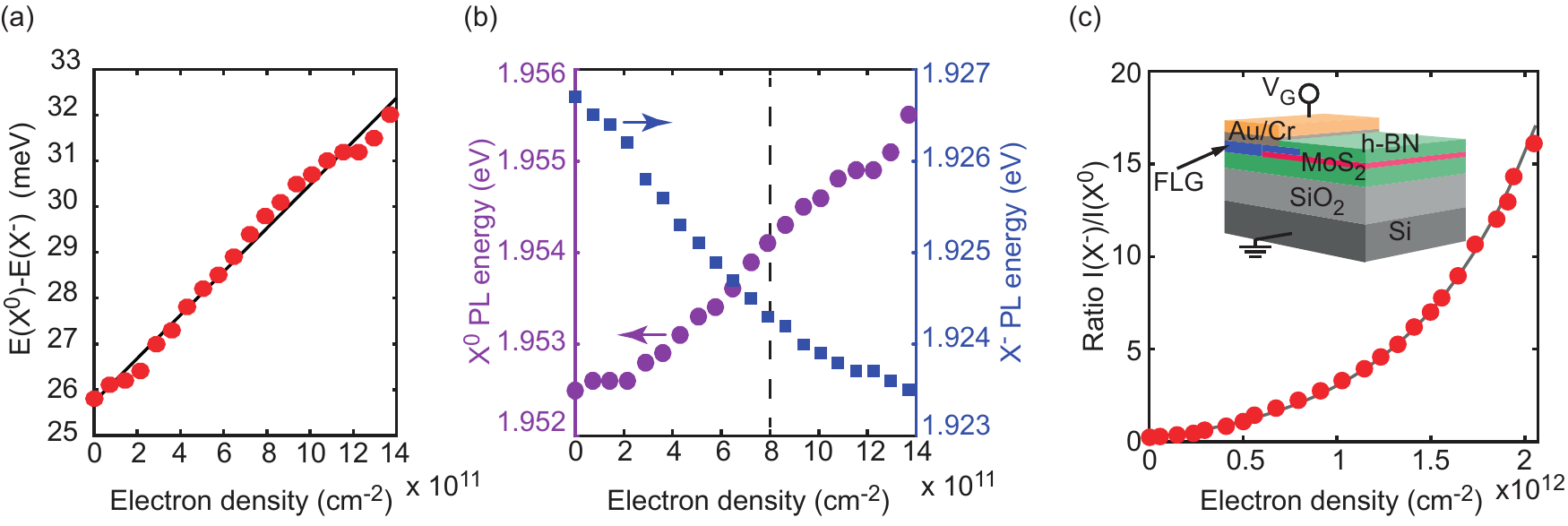}
\caption{Directly contacted MoS$_2$ device (reference sample). (a) Separation between the emission energy of the neutral $E(\tr{X}^0)$ and negatively charged $E(\tr{X}^-)$ excitons as a function of the electron density in the MoS$_2$ monolayer. (b) Variation of the photoluminescence energy of X$^0$ (purple circles, left axis) and X$^-$ (blue squares, right axis) as a function of the electron density. The vertical dashed line indicates the electron density at which the measurements in Fig.~\ref{Fig4} on the main sample were carried out. (c) Ratio between the integrated intensity of the X$^-$ and X$^0$ photoluminescence features as a function of the electron density. Inset: three-dimensional schematic view of the device. The MoS$_2$ is contacted at one side by a few-layer graphene electrode.}
\label{Fig3}
\end{center}
\end{figure*}

Figure~\ref{Fig2}(a) shows typical PL spectra recorded over a voltage loop as illustrated in Fig.~\ref{Fig2}(c): V$_\tr{G}$ varies from the initial state (S$_1$) at $-85$ V to $+85$ V (S$_3$) via S$_2$ (0~V) and then back to S$_1$ via S$_4$ (0~V). Two prominent features can be clearly identified (see Fig.~\ref{Fig2}(b)): a low-energy peak near $1.92~\tr{eV}$ and a high-energy peak near $1.95~\tr{eV}$ attributed to the negatively charged X$^{-}$ and the neutral X$^{0}$ A-excitons~\cite{Mak2013,Cadiz2017}, respectively. The emission energies of X$^{-}$ and X$^{0}$ change with $V_\tr{G}$, as seen in Fig.~\ref{Fig2}(d) where the X$^{0}$ energy has been plotted. However, as demonstrated in the colormap in Fig.~\ref{Fig2}(a), the intensities of the X$^{-}$ and X$^{0}$ features also vary with $V_\tr{G}$. The ratio between the integrated PL intensities of X$^{-}$ and X$^{0}$, $I(\tr{X}^-)/I(\tr{X}^0)$ (Fig.~\ref{Fig2}(e)), cannot be explained by the QCSE as it depends on the gate voltage sweep direction. Instead, the change in relative intensity arises from a change in the electron density~\cite{Manassen1996}.

In order to monitor the carrier density and its relation to $I(\tr{X}^-)/I(\tr{X}^0)$, a reference sample consisting of an encapsulated yet contacted MoS$_2$ monolayer was fabricated as sketched in the inset to Fig.~\ref{Fig3}(c). In this case, the MoS$_2$ layer is directly contacted by a few-layer graphene sheet. This is a capacitive device and as such the electron density in the sample is expected to change linearly with the applied gate voltage~\cite{Mak2013}. This expectation was confirmed experimentally by measuring the energetic separation between the X$^{0}$ and X$^{-}$ features in the PL spectra: we find a linear dependence of the X$^{0}$ and X$^{-}$ energy separation with gate voltage (see Fig.~\ref{Fig3}(a)). At low electron densities, the energetic separation between X$^0$ and X$^-$ scales linearly with the Fermi level, as ionization of X$^-$ requires that an electron is moved up to the Fermi level~\cite{Hawrylak1991,Mak2013}. Given the linear dependence of Fermi energy on electron density for a two-dimensional system, the PL itself demonstrates that the reference sample charges as a capacitive device (with a capacitance $\approx 12~\tr{nF}~\tr{cm}^{-2}$). It is noteworthy that the X$^{0}$ and X$^{-}$ emission energies show opposite dependences on the electron density (Fig.~\ref{Fig3}(b)):  X$^{0}$ blue-shifts while  X$^{-}$ red-shifts with increasing electron density~\cite{Chernikov2015}. As in the main sample, the reference sample shows hysteresis effects on ramping the voltage up and down due to photodoping from the surrounding h-BN~\cite{Ju2014,Epping2016}. The voltage at which the electron density is close to zero changes depending on the history of the device. However, we find a robust relationship between the intensity ratio $I(\tr{X}^-)/I(\tr{X}^0)$ and the X$^{0}$, X$^{-}$ splitting, equivalently the electron density.  Fig.~\ref{Fig3}(c) plots $I(\tr{X}^-)/I(\tr{X}^0)$ as a function of the electron density extracted from the PL spectra recorded at various gate voltages on the reference sample. This means that the ratio $I(\tr{X}^-)/I(\tr{X}^0)$ can be used as a measure of the electron density. The monotonic increase of this ratio with the electron density can be well described by a phenomenological exponential fit. This is used here as a calibration curve to evaluate the electron density in the main sample from the $I(\tr{X}^-)/I(\tr{X}^0)$ ratio.

Using the density calibration from the reference sample, the variation of the intensity ratio $I(\tr{X}^-)/I(\tr{X}^0)$ along the voltage loop displayed in Fig.~\ref{Fig2}(e) indicates a total variation of the electron density of $\sim 10^{12}~\tr{cm}^{-2}$ in the main sample. This change in electron density when applying a gate voltage might arise as a combined consequence of photodoping effects~\cite{Ju2014,Epping2016}, tunneling~\cite{Choi2013} from the FLG top gate through the insulating h-BN top layer, and charge trapping~\cite{Wang2010} at the SiO$_2$/h-BN interface. In order to isolate the QCSE contribution to the exciton energy, it is important to identify regions where the gate voltage can be swept without changing the ratio $I(\tr{X}^-)/I(\tr{X}^0)$. Inspection of the $I(\tr{X}^-)/I(\tr{X}^0)$ data in Fig.~\ref{Fig2}(e) shows that there are no significant changes in MoS$_2$ electron density around S$_3$ and S$_4$. These two regions are therefore good candidates for measuring the QCSE in MoS$_2$.

Between S$_2$ and S$_3$, the ratio $I(\tr{X}^-)/I(\tr{X}^0)$ is small and corresponds to a region where MoS$_2$ has a low electron density ($\lesssim 7\times 10^{10}~\tr{cm}^{-2}$). In this region, the X$^-$ signal is weak and evaluation of the ratio $I(\tr{X}^-)/I(\tr{X}^0)$ becomes unreliable. It is therefore difficult to attest that the Fermi level in this region remains absolutely constant. Moreover, it is in this range of Fermi energy that photo-induced doping from the h-BN layers occurs leaving charged defects in the h-BN that potentially induce electric field screening~\cite{Ju2014}. The region around S$_3$ is therefore problematic with regards to the QCSE. The region around S$_4$, between $+30$~V and $-30$~V, exhibits a stronger X$^-$ feature and the ratio $I(\tr{X}^-)/I(\tr{X}^0)$ can therefore be reliably measured. From this ratio, the electron density is evaluated to be $8 \times 10^{11}~\tr{cm}^{-2}$ in this region using the calibration curve displayed in Fig.~\ref{Fig3}(c) and remains constant to within 5\%.

\begin{figure}[!t]
\begin{center}
\includegraphics[width=8.6cm]{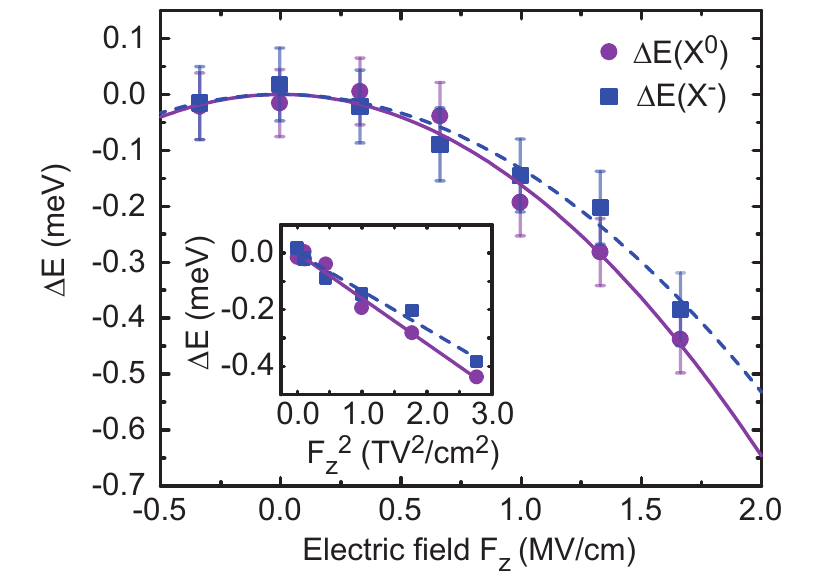}
\caption{Neutral X$^0$ (purple circles) and negatively charged X$^-$ (blue squares) Stark shift as a function of the applied electric field $F_z$, extracted from the measurements in Fig.~\ref{Fig2}. The solid purple and dashed blue lines are parabolic fits. The inset shows the same data points as a function of the squared electric field $F_z^2$ in order to highlight the quadratic dependence of the Stark shift.}
\label{Fig4}
\end{center}
\end{figure}

Fig.~\ref{Fig4} displays the change in X$^{-}$ and X$^{0}$ emission energies, $\Delta E(\tr{X}^-)$ and $\Delta E(\tr{X}^0)$ respectively, in the region around S$_4$. $F_z$ was determined by dividing $V_\tr{G}$ by the electrode-to-electrode distance of 300~nm and adding a constant built-in electric field of $0.66~\tr{MV/cm}$. This value was chosen such that $\Delta E(\tr{X}^-)$ and $\Delta E(\tr{X}^0)$ vanish at $F_z=0$, i.e.\ it is assumed that $\mu_z=0$. This built-in electric field arises from space charge within the layers of the heterostructure. Both $\Delta E(\tr{X}^-)$ and $\Delta E(\tr{X}^0)$ exhibit a quadratic dependence on $F_z$, equivalently a linear dependence on $F_z^2$, as shown in Fig.~\ref{Fig4}. This is the signature of a QCSE. We argue that the experiment reveals a QCSE and not a residual effect of any small changes in carrier density. First, for each point the measurement error in $I(\tr{X}^-)/I(\tr{X}^0)$ results in an uncertainty in the carrier density which leads to possible changes in $\Delta E(\tr{X}^-)$ and $\Delta E(\tr{X}^0)$ even without a QCSE. However, these changes (shown by the error bars in Fig.~\ref{Fig4}) are considerably smaller than the $\Delta E(\tr{X}^-)$ and $\Delta E(\tr{X}^0)$ values observed experimentally: the uncertainties in electron density cannot account for the shifts in X$^{-}$ and X$^{0}$ emission energies. Second, both X$^0$ and X$^{-}$ red-shift around S$_4$ yet a change in electron density would result in $\Delta E(\tr{X}^-)$ and $\Delta E(\tr{X}^0)$ values of opposite sign (see Fig.~\ref{Fig3}(b)). (Note that X$^0$ and X$^{-}$ are measured simultaneously.) From a fit to a second order polynomial, $\Delta E(\tr{X}^-)$ and $\Delta E(\tr{X}^0)$ versus $F_z$, excitonic polarizabilities of $\beta_z(\tr{X}^{-})= (6.4~\pm~0.9)\times 10^{-10}~\tr{D~m~V}^{-1}$ and $\beta_z(\tr{X}^{0})= (7.8~\pm~1.0)\times 10^{-10}~\tr{D~m~V}^{-1}$ are deduced at an electron density of $8 \times 10^{11}~\tr{cm}^{-2}$. These values are nearly one order of magnitude lower than the previously reported values~\cite{Klein2016}. 

Although the polarizabilities were measured at an electron density of $8 \times 10^{11}~\tr{cm}^{-2}$, $\beta_z(\tr{X}^{-})$ and $\beta_z(\tr{X}^{0})$ should remain constant for carrier densities lower than $\sim 10^{12}~\tr{cm}^{-2}$ where the conventional excitonic picture is valid~\cite{Sidler2016,Efimkin2017}. The transition to a many-body description occurs at higher electron densities than those used here. The slightly smaller polarizability of X$^{-}$ with respect to X$^{0}$ can be explained by Coulomb effects. Specifically, in the X$^{-}$ complex there is an additional decrease in the exciton binding energy with electric field. This is induced by the localization of the two electrons on one side of the monolayer, increasing the electron-electron repulsion~\cite{Shields1998}. The excitonic polarizabilities have been theoretically calculated with a finite barrier quantum well model~\cite{Pedersen2016}. Using barriers of 2.8~eV~\cite{Choi2013} for both electron and hole, a quantum well thickness of 0.65~nm and effective electron and hole masses of 0.35 bare electron mass~\cite{Mak2013}, an exciton polarizability of $7.5\times 10^{-10}~\tr{D~m~V}^{-1}$ is deduced which is in good agreement with the values reported here.

In conclusion, the QCSE of excitons has been extracted from photoluminescence measurements on a high quality MoS$_2$ monolayer embedded in a vdWh. As the electron density in the monolayer is observed to vary with electric field, a careful data analysis exploiting reference measurements on a directly contacted MoS$_2$ device was performed. Regions were identified in which the carrier density in the monolayer remains constant as the electric field is varied. Having ruled out any contribution of a changing electron density to the exciton energy shift, a QCSE was unambiguously identified. The small exciton polarizability is in line with theoretical computation~\cite{Pedersen2016}. The maximum QCSE achieved here corresponds to just half the homogeneous linewidth despite the fact that large eletric fields were applied. The insensitivity of the exciton to an electric field in MoS$_2$ has profound implications on its optical properties. On the one hand, we believe that the minute QCSE renders the exciton energy insensitive to charge noise. This, along with the super-fast radiative decay, explains the observation of optical linewidths close to the homogeneous limit in MoS$_2$ vdWhs~\cite{Cadiz2017}. On the other hand, electrical control of the exciton based on the QCSE would require larger polarizabilities or a non-zero dipole moment as observed in heterobilayers for instance~\cite{Rivera2015}. The methodology used here to determine the QCSE in MoS$_2$ can be used also in other semiconducting monolayers, where similar values of the polarizability should be obtained owing to the extreme out-of-plane confinement of both electrons and holes.

\begin{acknowledgements}
We thank M. Munsch for fruitful discussions. We acknowledge financial support from SNF Project 200020\_175748, Swiss Nanoscience Institute, QCQT PhD School, NCCR QSIT and Graphene Flagship. Growth of hexagonal boron nitride crystals was supported by the Elemental Strategy Initiative conducted by the MEXT, Japan and JSPS KAKENHI Grant Numbers JP15K21722.
\end{acknowledgements}

\section{Methods}

\textbf{\textit{Device fabrication}}
Van der Waals heterostructures were fabricated by stacking two-dimensional materials via a dry-transfer technique~\cite{Zomer2014}. All layers were mechanically exfoliated from bulk crystals (natural MoS$_2$ crystal from SPI Supplies, synthetic h-BN~\cite{Taniguchi2007} and natural graphite from NGS Naturgraphit). MoS$_2$ monolayers were treated by a bis(tri-fluoromethane)sulfonimide (TFSI) solution following Ref.~\onlinecite{Amani2015} before full encapsulation between h-BN layers. Few-layer graphene was employed as a top transparent electrode or as a contact electrode to MoS$_2$~\cite{Yu2014}. Metal contacts to FLG were patterned by electron-beam lithography and subsequent metal deposition of Au (45~nm)/Cr (5~nm). The flake thickness of each layer was characterized by a combination of optical contrast, atomic force microscopy, PL and Raman spectroscopy. The data shown in this Letter were measured on a device consisting of SiO$_2$ (300~nm)/h-BN (5.4~nm)/MoS$_2$ (0.65~nm)/h-BN (12~nm)/FLG (17~nm).\\
\textbf{\textit{Photoluminescence measurements}}
Photoluminescence spectroscopy was performed in a liquid He bath cryostat using a home-built confocal microscope setup. The main sample and the reference sample were optically excited using a linearly polarized diode laser at photon energy 2.32~eV (wavelength 535~nm) and a HeNe laser at photon energy 2.09~eV (wavelength 594~nm) with an intensity below $2~\tr{kW~cm}^{-2}$, respectively. The collected light was dispersed onto a charged-coupled device array by a single monochromator equipped with a 1500 grooves/mm grating.

%\bibliography{H2D_2017_10}


\end{document}